\documentclass[manuscript,longabstract]{aastex}
\usepackage{latexsym,wasysym}

\usepackage{latexsym}

\def\al26{\mbox{$^{26}$Al}}
\def\ca41{\mbox{$^{41}$Ca}}
\def\fe60{\mbox{$^{60}$Fe}}
\def\osixteen{\mbox{$^{16}$O}}
\def\oseventeen{\mbox{$^{17}$O}}
\def\oeighteen{\mbox{$^{18}$O}}
\def\msun{\mbox{M$_{\odot}$}}

\def\delseventeen{\mbox{$\delta^{17}$O}}
\def\deleighteen{\mbox{$\delta^{18}$O}}
\def\ctwelve{\mbox{$^{12}$C}}
\def\cthirteen{\mbox{$^{13}$C}}

\slugcomment{In press in Astrophysical Journal Letters}

\shorttitle{Oxygen isotopic composition of the Solar System}
\shortauthors{Gaidos et al.}

\begin{document}

\title{On the oxygen isotopic composition of the Solar System}

\author{Eric Gaidos}\affil{Department of Geology and Geophysics,
University of Hawaii, Honolulu, HI 96822, USA}
\email{gaidos@hawaii.edu}

\and

\author{Alexander N. Krot and Gary R. Huss} \affil{Hawaii Institute of Geophysics and Planetology,
University of Hawaii, Honolulu, HI, 96822, USA}
\email{sasha@higp.hawaii.edu,huss@higp.hawaii.edu}

\begin{abstract}

  The \oeighteen/\oseventeen~ratio of the Solar System is 5.2 while
  that of the interstellar medium (ISM) and young stellar objects is
  $\sim$4.  This difference cannot be explained by pollution of the
  Sun's natal molecular cloud by \oeighteen-rich supernova ejecta
  because (1) the necessary B-star progenitors live longer than the
  duration of star formation in molecular clouds; (2) the delivery of
  ejecta gas is too inefficient and the amount of dust in supernova
  ejecta is too small compared to the required pollution (2\% of total
  mass or $\sim$20\% of oxygen); and (3) the predicted amounts of
  concomitant short-lived radionuclides (SLRs) conflicts with the
  abundances of \al26 and \ca41 in the early Solar System.  Proposals
  for the introduction of \oeighteen-rich material must also be
  consistent with any explanation for the origin of the observed
  slope-one relationship between \oseventeen/\osixteen~and
  \oeighteen/\osixteen~in the high-temperature components of primitive
  meteorites.  The difference in \oeighteen/\oseventeen~ratios can be
  explained by enrichment of the ISM by the \oseventeen-rich winds of
  asymptotic giant branch (AGB) stars, the sequestration of
  comparatively \oeighteen-rich gas from star-forming regions into
  long-lived, low-mass stars, and a monotonic decrease in the
  \oeighteen/\oseventeen~ratio of interstellar gas.  At plausible
  rates of star formation and gas infall, Galactic chemical evolution
  does {\it not} follow a slope-one line in an three-isotope plot, but
  instead moves along a steeper trajectory towards an \oseventeen-rich
  state.  Evolution of the ISM and star-forming gas by AGB winds also
  explains the difference in the carbon isotope ratios of the Solar
  System and ISM.

\end{abstract}

\keywords{astrochemistry --- solar system: formation --- ISM:
evolution --- Galaxy: evolution --- stars: AGB and post-AGB}

\section{Was the protosolar cloud polluted by a supernova?}

Abundance ratios of the three stable isotopes of oxygen
(\osixteen,\oseventeen,\oeighteen) in meteorites and planetary samples
are used to infer the existence of reservoirs and processes in the
early Solar System.  Each of these isotopes was produced in a
different mass-distribution of stars, and the isotopic composition of
the Solar System is the cumulative result of $\sim$9 billion years
(Gyr) of previous Galactic stellar nucleosynthesis.  \osixteen, a
primary isotope, is principally formed during He-burning in massive
stars, and added to the ISM by core-collapse supernovae (Type II SNe).
\oseventeen~and \oeighteen~are secondary isotopes: The first is
produced by reaction of light nuclei during the CNO cycle; the second
by nitrogen reactions during He burning.  Both of the heavier isotopes
are ejected by Type II SNe.  \oseventeen~is also produced by the
intermediate-mass progenitors of asymptotic giant branch (AGB) stars.
The yield scales with the abundance of \osixteen~\citep{Meyer08}.
Thus, in a plot of \oseventeen/\osixteen~vs. \oeighteen/\osixteen,
Galactic chemical evolution (GCE) is expected to proceed along a
slope-one line \citep{Timmes95}.  Comparisons of the Sun with gas and
other objects in the Milky Way should reveal such a trend, with
younger objects being relatively \osixteen~poor.

Observations to date do not fulfil this expectation.  Measurements of
primordial Solar System materials, including the solar wind returned
by {\it Genesis} spacecraft \citep{McKeegan09}, fall along a slope-one
line that has a high \oeighteen/\oseventeen~(but not high \osixteen)
compared to the much younger ISM and newly-formed stars
\citep{Wannier80,Wilson94,Wouterloot08}.  One proposed explanation is
that the Solar System formed from molecular gas polluted by
\oeighteen-rich gas from massive stars
\citep{Olive82,Henkel93,Young09}.  SLRs such as \al26, \ca41 and \fe60
in the early Solar System are interpreted as evidence that
contamination did occur \citep{Goswami05}.  However, the oxygen
isotopic composition of the Solar System falls along the same trend
inferred for the initial compositions of AGB stars that formed before
the Sun and produced the presolar oxide grains in primitive meteorites
\citep{Nittler09}.  Furthermore, to explain the oxygen isotopic
composition of the Solar System, (1) the delivery of SN ejecta must
have occurred before cessation of star formation in the cloud, (2) it
must have been efficient enough to produce the observed offset, (3) it
must be quantitatively consistent with the abundance of SLRs, and (4)
it must be compatible with the observed slope-one dispersion of
primitive Solar System materials and the solar wind in a three-isotope
plot (\delseventeen~vs. \deleighteen), if this dispersion is primary
(see below).  We argue that the SN pollution hypothesis fails all four
tests.

First, stellar nucleosynthesis calculations show that {\it only}
  B-star progenitors (8-15~\msun) have SN ejecta with
  \oeighteen/\oseventeen~ratios substantially higher than the current
  ISM \citep{Meyer08} and are plausible sources of pollution.
  However, these stars have main-sequence lifetimes $>$15~Myr, longer
  than the typical duration of star formation in clouds
  \citep{Williams00,Lada03}, and thus low-mass stars
  forming from the same cloud are unlikely to be polluted.  The
  probability of an unaffiliated B star polluting a given cloud during
  its lifetime is approximately $R \tau f$, where $R \sim 2 \times
  10^4$~Myr$^{-1}$ is the Type II SN rate in the Milky Way
  \citep{Dragicevich99} (mostly from B star progenitors), $\tau \sim
  10$ Myr is the mean cloud lifetime \citep{Lada03}, and $f \sim
  10^{-8}$ is the volume filling factor of a 10~pc cloud in the
  Galactic disk.  The probability is $\sim$0.2\%, comparable to that
  of an encounter with an AGB star \citep{Kastner94}.

Second, mixing between hot, tenuous SN ejecta and cooler, denser
molecular gas is inefficient.  A 3-dimensional simulation produced
mixing of 50 ppm \citep{Boss08}, only 2.5\% of what is required.
Refractory grains of Al$_2$O$_3$ can carry \al26 into a cloud
\citep{Gaidos09}, but can accommodate only 0.9\% of all oxygen
\citep{Lodders03}.  SN are predicted to be copious sources of dust,
but no more than $10^{-5}$-$10^{-4}$~\msun~has been detected per event
\citep{Meikle07}.  The scarcity of \osixteen-rich presolar grains of
plausible SN origin \citep{Hoppe00} also supports inefficient SN dust
production and transfer to star-forming regions.

Third, SLRs such as \al26, \ca41, and \fe60, with half-lifes of 0.72,
0.1~Myr, and 2.6~Myr \citep{Rugel09}, respectively, accompanied SN
ejecta and Wolf-Rayet winds from massive stars, and SLR abundances and
any oxygen isotopic shift should be related \citep{Gounelle07}.
Previous work has shown that introduction of exogenous SLRs would
likely have occurred during the giant molecular cloud phase
\citep{Gaidos09}.  \al26 was homogeneously distributed in
the protosolar nebula with the ratio \al26/$^{27}$Al $\approx 5 \times
10^{-5}$ \citep{Thrane06,Villeneuve09}.  The allowed fraction of SN
ejecta depends on the interval of free decay of \al26 between
injection and the formation of refractory calcium-aluminum-rich
inclusions (CAIs, taken to represent time zero).  The abundance of
\ca41 constrains the interval of free decay to $\sim$0.3~Myr, and
Monte Carlo calculations show that the Solar System most likely formed
from a cloud polluted to 0.3\% by a generation of stars that formed
4.5 Myr earlier \citep{Gaidos09}.  Only progenitors more massive than
50 \msun~could pollute, but such stars are depleted in both
\oseventeen~and \oeighteen~with respect to the Solar System.  Limongi
\& Chieffi (2003) predict that the ejecta from a 35~\msun~progenitor
contains 5, $1.7 \times 10^{-5}$, and $5 \times 10^{-5}$ \msun~of
\osixteen, \oseventeen, and \oeighteen, respectively.  [See Woosley \&
  Weaver (1995) for similar figures.]  This \osixteen-rich material
leaves the \oeighteen/\oseventeen~ratio of the cloud essentially
unchanged.  Ejecta from higher-mass progenitors is expected to be even
more depleted in \oseventeen~and \oeighteen.

The high-temperature components (chondrules and CAIs) of unaltered
chondritic meteorites plot along a slope-one line in a three-isotope
oxygen diagram \citep{Yurimoto07}\footnote{Only rare FUN
(Fractionation and Unidentified Nuclear effects) and F (Fractionation)
CAIs show evidence for extensive mass-dependent isotope fractionation
due to melt evaporation and plot along lines with slope of ~0.52
\citep{Krot09}}. The dispersion is usually explained by models in
which (1) the initial oxygen isotopic compositions of both solids and
gas were identical and equal to the solar wind value returned by {\it
Genesis}; and (2) the slope-one line is a consequence of mixing with
an \osixteen-poor reservoir generated in the protoplanetary disk by
``self-shielding'' of CO isotopomers from ultraviolet dissociation,
e.g. Lyons \& Young (2005).  However, these assumptions have been
challenged by Krot et al. (2009), who concluded that (1) primordial
(thermally unprocessed) solids in the Solar System were already
\osixteen-depleted relative to solar nebula gas; and (2) CO
self-shielding had only a small effect on the isotopic chemistry of
dust in the Solar System.  In this scenario, the slope-one line was
produced by mixing of {\it relict}, isotopically distinct reservoirs,
i.e. gas and dust of different ages \citep{Dwek06}.  Any contamination
by \oeighteen-rich ejecta would have to be contrived such that the
final slope-one line was preserved.

\section{The Solar System in the context of Galactic chemical evolution}

The isotopic composition of the Solar System can be explained in terms
of GCE that has deviated from the canonical slope-one line at least
since the Sun formed 4.6 Gyr ago.  A flat age-metallicity relation
\citep{Holmberg07} and the deficit of metal-poor stars in the solar
neighborhood \citep{Caimmi08} indicate that the metallicity of
star-forming gas has evolved little over the past $\sim$10~Gyr,
presumably because of the infall of metal-poor gas and the
sequestration of metals in low-mass stars \citep{Colavitti08}.  This
means that the recent evolution of the \oseventeen/\osixteen~and
\oeighteen/\osixteen~ratios in the Milky Way depended not on
\osixteen, but on the relative contributions by stars of different
masses to each isotope, and the time-variation caused by the strong
dependence of main sequence lifetime on stellar mass.

The ISM is neither a closed system nor chemically and isotopically
homogeneous.  Gas and dust move through different phases (molecular
clouds, neutral ISM, HII regions), and stars with different masses add
or remove isotopes at different times depending on their main sequence
life. A GCE model should include: (1) infall of metal-poor,
\osixteen-rich gas; (2) massive progenitors of SN (and Wolf-Rayet
stars) which begin ejecting mass 1-3~Myr after the formation of their
progenitors \citep{Limongi03} and can pollute their natal molecular
cloud \citep{Gaidos09}; (3) stars with initial masses of
1.5-8 \msun~which move onto the red giant and asymptotic branches
$>$70~Myr after their formation, long after dispersal of the molecular
cloud, eject mass into low-density ISM phases, and preferentially
enrich them in \oseventeen~with respect to star-forming
regions\footnote{One manifestation of this is the appearance of
\oseventeen-rich presolar grains in primitive meteorites
\citep{Nittler97,Nittler09}.}; and (4) stars with $<$1~\msun~which
remain on the main sequence for $>$12~Gyr and are a sink for all
isotopes.  The sequestration of relatively \oeighteen-rich gas from
star-forming regions will leave the ISM enriched in \oseventeen.  We
propose that the cumulative effect is GCE with a stagnant
\oeighteen/\osixteen~ratio but a decreasing
\oeighteen/\oseventeen~ratio, and that the difference between the Sun
(and all stars of its age) and gas in the present Milky Way is a
result of this gradual change.

We calculated the evolution of the oxygen isotopic composition in the
vicinity of the Sun's formation and present location using a two-box
model of GCE.  One box represents the low-density phases of the ISM
(hereafter referred to as the ISM), which receives metal-poor
infalling gas, gas ejected from star-forming regions, winds from AGB
stars, and some SN ejecta.  The second box represents star-forming
regions (molecular clouds).  It receives gas from the ISM, returning
(most of) it after disruption of the clouds, and the difference is
incorporated into stars.  Three types of stars are considered:
low-mass stars ($<1$ \msun) which only sequester mass;
intermediate-mass progenitors of AGB stars, and high-mass ($>8$~\msun)
SN progenitors.  SNe eject mass into either molecular clouds or the
ISM, depending on their main-sequence life relative to the cloud
lifetime (10~Myr); AGB stars pollute only the ISM.  The mass flux from
the ISM into molecular clouds is calculated using a fixed ISM
residence time of 280~Myr, a value calculated from a total gas surface
density of 14~\msun~pc$^{-2}$ in the present solar neighborhood
\citep{Naab06}, a star formation efficiency of 10\%
\citep{Williams00,Lada03}, and a local star-formation rate of
$5\times10^{-3}$~\msun~pc$^{-2}$~Myr$^{-1}$.  This last value is based
on a SN rate of 0.04~yr$^{-1}$ \citep{Dragicevich99} and the abundance
of \al26 in the Milky Way \citep{Diehl06}.  This residence time is
consistent with previous estimates \citep{Tenorio00} and the
interstellar residence times of presolar grains \citep{Heck09}.  The
return flow of gas through HII regions and the disruption of molecular
clouds by SN assumes a cloud residence time of 10~Myr
\citep{Williams00,Lada03}.  We relate star-formation rate to total gas
density by a Schmidt-Kennicut law with an index of 1.45
\citep{Fuchs09}.  We adopt a power-law stellar initial mass function
with index of 2.35 for $>1$~\msun~and take the fraction of stellar
mass in lower-mass stars to be 0.42.

The present gas infall rate is poorly constrained by direct
observations, but rates have been inferred from the distribution of
stellar metallicities.  We assume a constant rate based on a recent
analysis of G- and K-star metallicities \citep{Casuso04}.  Moreover,
the adopted relationship between star-formation and gas surface
density, and the current {\it total} mass density in the solar
neighborhood [$\sim$50~\msun~pc$^{-2}$ \citep{Naab06}] restrict the
range of possible infall rates in our model.  A rate of $3.5 \times
10^{-3}$~\msun~pc$^{-2}$~Myr$^{-1}$ successfully reproduces the
present surface densities of gas, stars, and stellar remnants [14, 35,
and 3 \msun~pc$^{-2}$, respectively \citep{Naab06}].  The infalling
gas is assumed to contain 10\% solar \osixteen~and no \oseventeen~ or
\oeighteen.

Nucleosynthetic yields are taken from Woosley \& Weaver (1995) for SN
and from Boothroyd \& Sackmann (1999) for AGB stars.  We do not
consider stars more massive than 40 \msun, for which detailed yields
have yet to be published.  These stars are thought to contribute
negligible \oseventeen~or \oeighteen.  Dependence on metallicity is
determined by linear interpolations between solar and single sub-solar
values.  The \oseventeen~yield from massive stars might have to be
adjusted significantly downwards because of revised values for the
proton capture reaction rates \oseventeen(p,$\alpha$)$^{14}$N and
\oseventeen(p,$\alpha$)$^{18}$F \citep{Meyer08}.  Removing the
contribution from massive stars to \oseventeen~altogether results in a
model \oseventeen/\osixteen~ratio in star-forming regions 4.6 Gyr ago
that is within 15\% of the solar value.  With unadjusted
\oeighteen~yields, the model predicts an \oeighteen/\osixteen~ratio
that is within 7\% of the solar value.  We are unable to
simultaneously reproduce the surface densities of stars and gas, {\it
and} the absolute oxygen abundance of star-forming clouds 4.6 Gyr ago
(i.e. the solar abundance).  One solution, besides tuning yields, is
start the assembly of the disk at the Sun's location only 10~Gyr ago,
i.e. 3.5~Gyr after the formation of the Galactic center
\citep{Naab06}.  Another is to allow a small fraction of the
oxygen-enriched gas from star-forming regions to permanently escape
the disk and be replaced with metal-poor gas.

Figure \ref{fig.evo} shows the predicted evolution of the oxygen
isotope ratios in both the ISM (diamonds) and star-forming regions
(squares) with respect to values in star-forming regions 4.6 Gyr ago
(taken to represent the solar value).  GCE is towards a
\osixteen-poor, but relatively \oseventeen-rich condition: it does
{\it not} follow a slope-one line.  In our model, the difference in
\oeighteen/\oseventeen~between the ISM and star-forming regions is due
only in small part to contamination of clouds by SN, because only
progenitors of 18-40 \msun~are sufficiently short-lived and these
contribute little \oeighteen.  Instead, it is mostly due to
contamination of the ISM with \oseventeen~from AGB stars over the
$\sim$280~Myr it takes for the two reservoirs to exchange.

Episodic gas infall or star formation produces yet more dramatic
departures from a slope-one trajectory.  Cosmological models predict
stochastic time variation associated with infalling dark matter halos
\citep{Colavitti08}, and accompanied by elevated rates of star
formation.  If the Sun formed during a $\sim$1~Gyr episode of
increased star formation (twice the background rate), it partakes of
an enhanced \oeighteen-rich contribution from massive stars (Figure
\ref{fig.episodicsf}) before winds from AGB stars return the ISM to
its previous isotopic trajectory.  In a sense, this {\it is} SN
pollution, but on a Galactic scale.

\section{Discussion}

\begin{figure}
\epsscale{1.0} 
\plotone{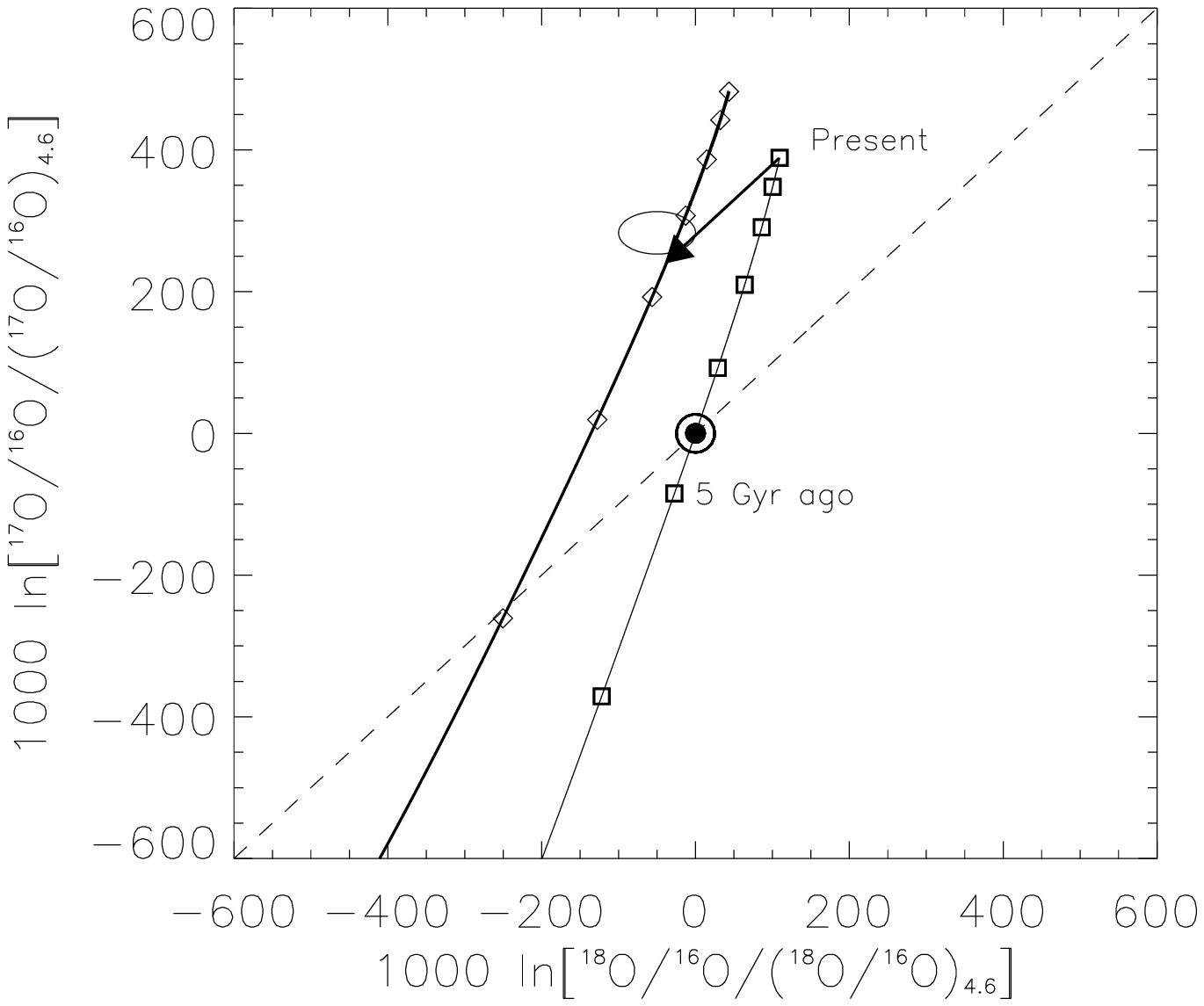}
\caption{Predicted evolution over 10~Gyr of the three oxygen isotope
in the ISM (diamonds) and star-forming regions (squares) at 1~Gyr
intervals, terminating in the upper right at the present day.  Units
are $1000\times$ the natural log of ratios with the value 4.6 Gyr ago
when the Sun formed.  The open ellipse is the composition of the ISM
relative to the Solar System \citep{Wouterloot08}.  The dashed line is
the standard slope-one GCE trajectory.  The arrow is the mixing line
produced by adding 20\% gas with 10\% solar \osixteen~and zero
\oseventeen~and \oeighteen~to present-day star-forming regions.
\label{fig.evo}}
\end{figure}

\begin{figure}
\epsscale{1.0} 
\plotone{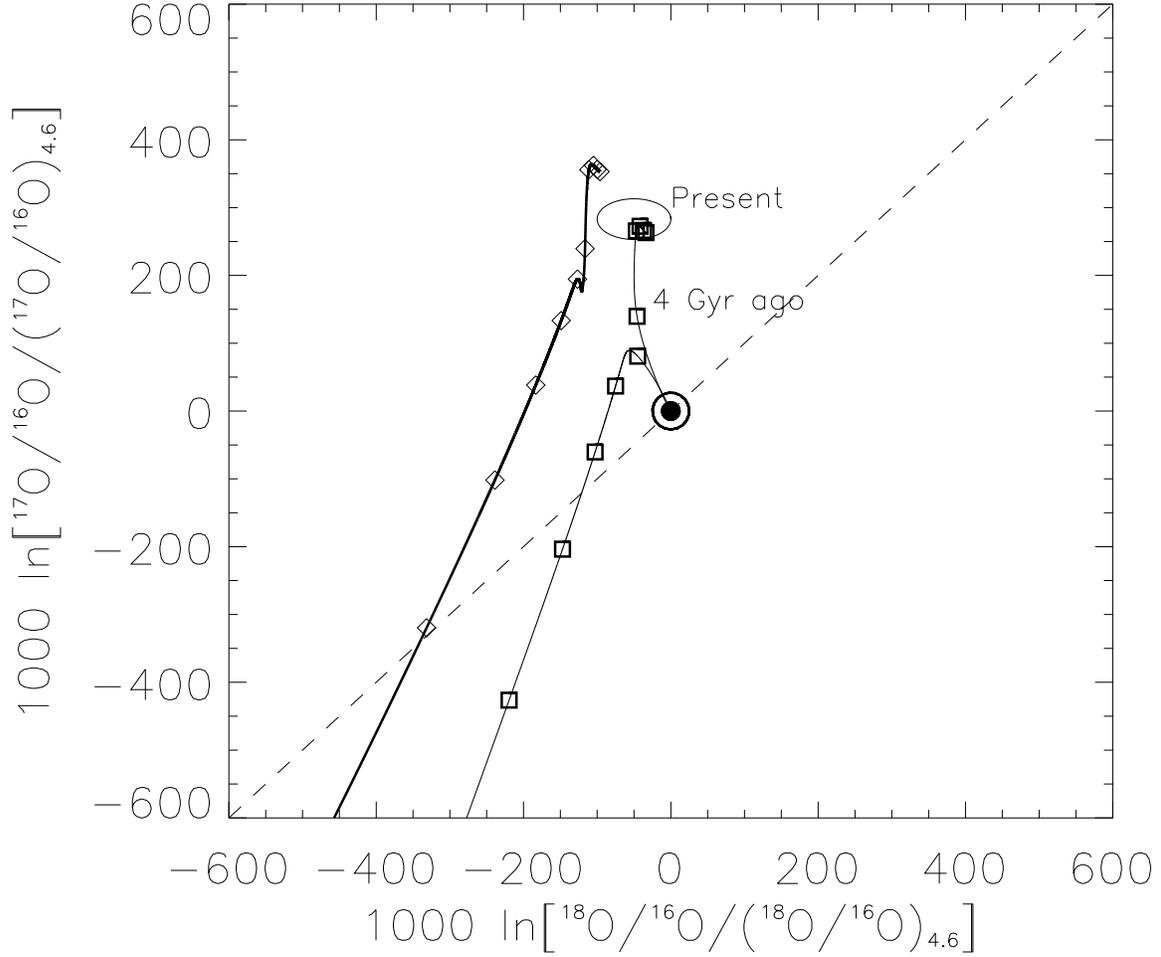}
\caption{Same as Figure 1, except with a period of elevated
($2\times$) star formation coinciding with the birth of the Sun.  The
event is modeled as a gaussian with a standard deviation of 250~Myr.
During the event the \oeighteen/\oseventeen~ratio increases as more
massive stars eject \oeighteen~into the ISM; more slowly-evolving AGB
stars add \oseventeen-rich gas later, returning the ISM and
star-forming regions to their former paths.
\label{fig.episodicsf}}
\end{figure}

Our model of GCE in the solar neighborhood produces a super-unity
slope in a three-oxygen isotope plot, and can explain the high
\oeighteen/\oseventeen~ratio of the Sun with respect to the ISM.  The
absence of this phenomenon in previous models \citep{Prantzos96} may
be attributable to the assumption that stars form from a gas of mean
disk composition, rather than one to which AGB stars of the same
generation have yet to contribute.  Our model also predicts a
difference between the composition of the ISM and star-forming
regions: such a difference has not been observed \citep{Young09} but
{\it measurements of oxygen isotopes are currently possible only in
molecular (i.e. CO-containing) gas} and the composition of the
low-density phases of the ISM is unknown.  The composition of young
stellar objects also appears to be dispersed along a slope-one line in
a three oxygen isotope plot.  This effect could be produced by
heterogeneous mixing of Galactic disk gas with infalling, metal-poor
gas lacking \oseventeen~and \oeighteen~(Figures \ref{fig.evo} and
\ref{fig.episodicsf}).

A purported Galactocentric gradient in oxygen isotopes (increasing
\osixteen~with radius) \citep{Wilson94} is cited as support for
standard GCE: However, recent data do not exhibit a significant trend
\citep{Polehampton05}, and while the \oeighteen/\osixteen~ratio at the
Galactic center may be higher than disk values, the former may have
been previously overestimated \citep{Wouterloot08}.  Instead,
Wouterlout et al. (2008) report an increasing
\oeighteen/\oseventeen~ratio with Galactocentric radius and a lower
\oeighteen/\oseventeen~in the Large Magellanic Cloud (LMC) compared to
similarly metal-poor regions of the outer disk.  They explain both of
these observations as a result of the accumulation of \oseventeen~with
time, i.e. both the inner disk and LMC are older than the outer disk.
A corollary of this conclusion is that at any given Galactocentric
radius, younger objects (e.g., the ISM and newly-formed stars) will be
more \oseventeen-rich than older ones (e.g., the Solar System).

Our scenario predicts that most low mass stars of the Sun's age should
have a similar oxygen isotopic composition.  The SN contamination
scenario, in contrast predicts that the Sun lies at the extreme of a
distribution of \oeighteen/\oseventeen~ratios (corresponding to
different levels of contamination), something the observations so far
do not suggest.  While it is presently not possible to measure the
oxygen isotopic composition of main-sequence stars, other isotopes
such as those of Mg \citep{Yong04} or Si \citep{Tsuji94} might serve
as proxies.

\begin{figure}
\epsscale{1.0} \plotone{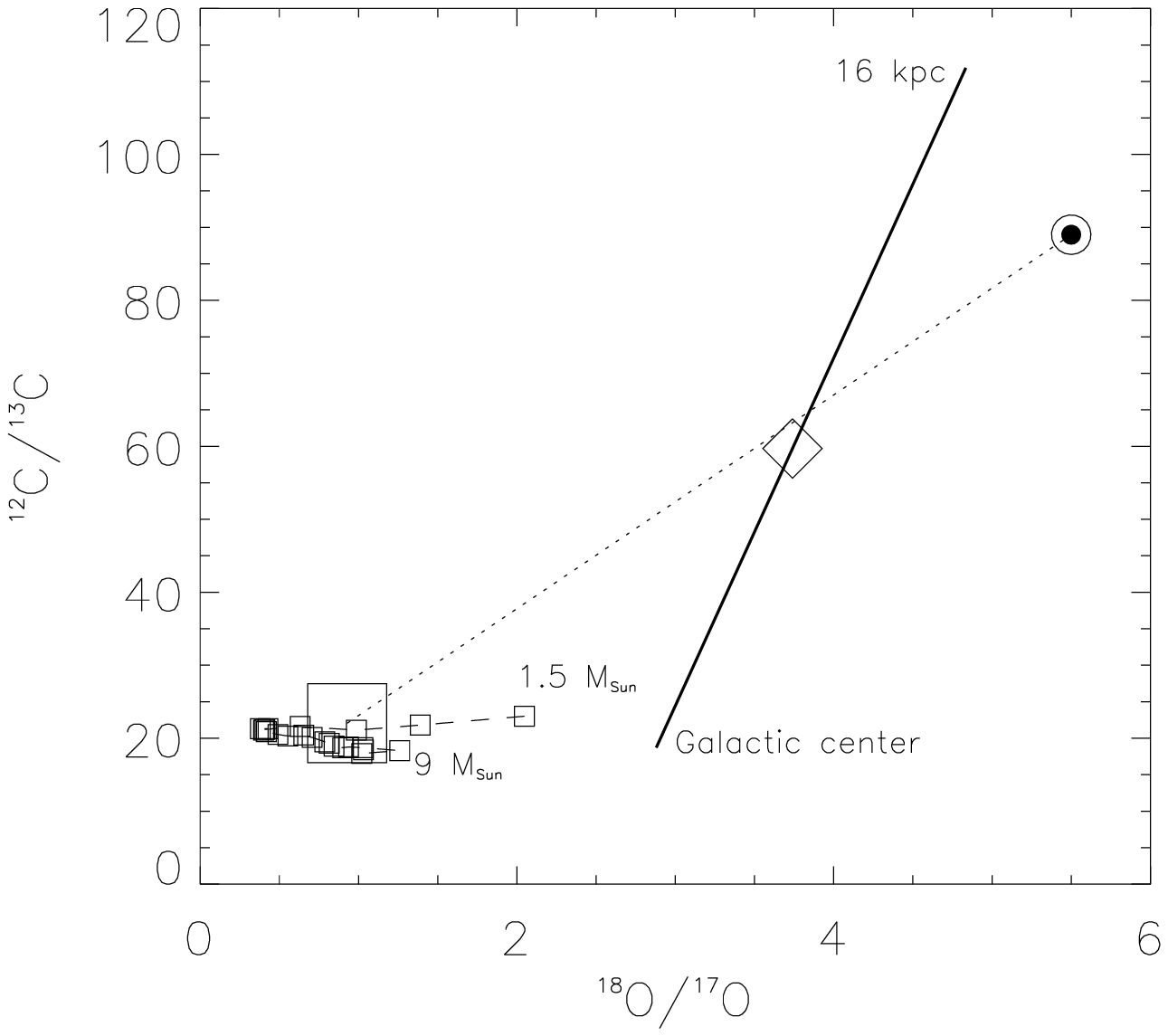}
\caption{\ctwelve/\cthirteen~ratio vs. \oeighteen/\oseventeen~ratio in
  the Milky Way: The solid line is the Galactocentric gradient in the
  ISM based on Milam et al. (2005) and Wouterlout et al. (2008).  The
  circle is the Solar System value, and the composition of
  1.5-9~\msun~AGB star envelopes after second dredge-up
  \citep{Boothroyd99} are small squares.  The large square is the
  IMF-integrated AGB ejecta composition assuming a Salpeter IMF and
  the ejecta mass functions of Boothroyd \& Sackmann (1999).  The
  dashed line is a hypothetical equal-metallicity mixing line between
  the mean AGB ejecta composition and interstellar gas 4.6 Gyr ago and
  at a Galactocentric radius of 6.6~kpc, represented by the Solar
  System \citep{Wielen96}.  The mixing line passes close to the
  present composition of the ISM at 6.6~kpc (diamond), indicating that
  it can be produced by addition of AGB ejecta to the ISM at that
  location 4.6 Gyr ago. \label{fig.co}}
\end{figure}

The \ctwelve/\cthirteen~ratio of the Solar System (89$\pm$1)
\citep{Wyckoff00,Clayton04b} is higher than the interstellar ratio in
the solar neighborhood ($76\pm7$) \citep{Stahl08}.  This difference
can be explained by the high \cthirteen/\ctwelve~ratio in winds from
red giant branch and AGB stars \citep{Timmes95,Milam05}, although
chemical inhomogeneities in the ISM may also be important
\citep{Stahl08}. The present-day C and O isotopic composition of the
ISM at the Galactocentric location of the Sun's birth is consistent
with mixing of average AGB winds with the ISM 4.6 Gyr ago (represented
by the Solar System) (Figure \ref{fig.co}).

Our arguments do not preclude {\it any} pollution of the Sun's parent
molecular cloud by massive stars, but demonstrate that the required
amount and kind of pollution is neither plausible nor necessary to
explain the \oeighteen/\oseventeen~ratio of the Solar System.  The
Solar System's initial inventory of \al26 from the ejecta of $M >
50$\msun~stars would have been accompanied by a shift of about
-65$\permil$ in both heavy isotopes.  If this \oseventeen- and
\oeighteen-depleted oxygen was delivered primarily as gas to the natal
molecular cloud, it would have established an offset between the
isotopic compositions of gas and dust in the cloud along the slope-one
line.  This, as well as mixing with infalling metal-poor gas, could
explain the slope-one dispersion in primitive meteorite components and
the solar wind \citep{Krot09}.

\acknowledgments

We thank Jonathan Williams and Ed Young for helpful discusssions, and
Roberto Gallino for a detailed review.  ANK and GRH acknowledge support
from NASA grants NNX07AI81IG and NNX08AG58G, respectively.

\clearpage

\end{document}